\documentclass{moriond}
\usepackage{amsmath}

\def\Journal#1#2#3#4{{#1} {\bf #2}, #3 (#4)}

\def\PRL{\em Phys. Rev. Lett.}
\def\PRD{{\em Phys. Rev.} D}

\def\al{\alpha}
\def\ab{\bar{\alpha}}
\def\ee{\end{equation}}
\def\bea{\begin{eqnarray}}
\def\eea{\end{eqnarray}}

\def\be{\beta}
\def\ga{\gamma}
\def\de{\delta}
\def\ep{\epsilon}
\def\ze{\zeta}
\def\et{\eta}

\def\ka{\kappa}
\def\la{\lambda}
\def\rh{\rho}

\def\si{\sigma}

\def\Ga{\Gamma}

\def\La{\Lambda}

\def\Om{\Omega}

\def\vth{\vartheta}

\def\vph{\varphi}

\def\cL{{\mathcal L}}
\def\cR{{\mathcal R}}

\def\mn{{\mu\nu}}
\def\ab{{\al\be}}

\def\pt#1{\phantom{#1}}
\def\prt{\partial}
\def\3g#1#2#3{^{(3)}\Ga^{#1}_{\pt{#1}#2#3}}

\newcommand{\Photo}{}

\begin{document}
\vspace*{4cm}
\title{Recent Progress in Gravity Tests of Spacetime Symmetries}

\author{ K. O'Neal-Ault }

\address{Department of Physics and Astronomy, Embry-Riddle Aeronautical University,\\
 Prescott, AZ, 86301, USA}

\maketitle\abstracts{We provide a brief overview into recent tests of gravity, focusing on its foundational spacetime symmetries. In particular, we work with an agnostic, effective field-theory framework, named the Standard-Model Extension, that allows for analysis of tests of such symmetries. There have been a wide range of experiments and theory developments that have helped contribute toward constraining terms within the framework, providing clues to a possible underlying, unified theory of physics.    
}
\section{Background}

The theory of General Relativity (GR) and the Standard Model (SM) of particle physics are both well tested, successful theories that describe physics for a wide range of distance scales in our universe. 
Yet it is still a challenge to find a satisfactory combination of the two theories. 
Both are expected to be descriptions for a low-energy limit of some fundamental theory of physics, that would appear at Planck scales of around $10^{19}$ GeV \cite{kost_2004}. Current experiments cannot directly probe such scales, thus instead we search for clues toward some unified physics; 
Lorentz symmetry (LV) and Charge-Parity-Time (CPT) symmetries are foundational to both GR and SM, 
and thus potentially detectable tiny violations could expose hints to new Plank-scale physics.

\subsection{An Effective Field Theory Framework}\label{subsec:prod}

We work with an effective field-theory framework named the Standard Model Extension (SME), that provides an agnostic, systematic approach to testing for spacetime symmetry breaking \cite{kost_2004,coll_kost_1998}.
The SME framework comprises of an action containing terms for GR and the SM, along with all possible additional, coordinate-independent terms that can allow for tests of Lorentz, CPT and diffeomorphism symmetries. 
These additional terms contract known fields operators with unknown coefficients, which control the degree of spacetime symmetry breaking. 
These coefficients are constrained through theoretical and experimental means. 
Each term also is classified by its mass dimension, which is in terms of natural units where $c=1$ and $\hbar=1$.

Numerous efforts in both experimental measurements and limits derived from theory have contributed toward the search for Lorentz and CPT violations within this model-independent framework. 
A collective source of decades of such work is the \textit{Data Tables for Lorentz and CPT Violation}, containing forty-six tables that cover eleven sectors of physics \cite{data_tables}. 
Within the gravity sector there are numerous tests of spacetime symmetries that were not covered within this proceeding, including atom interferometry, gyroscope experiment, short-range gravity, etc. 
Additionally, many publications contain mappings of current models of modified physics to terms within the SME framework, providing a means to constrain or rule out said models \cite{bk06,bc2018,jhf2021,km09}. 

\section{Gravity Tests}
This proceeding focuses solely on the gravity sector of the SME. 
Although there are numerous publications for tests of spacetime symmetry over the decades, this proceeding reasonably summarizes only a few \cite{data_tables}.

\subsection{Lunar Laser Ranging}

Lunar Laser Ranging provides about forty-five years of data, measuring the Earth-Moon distance via laser ranging \cite{LLR_1962}. 
The travel time for a laser beam originating from Earth, reflecting from the moon and returning to its source provides data precision on the order of micrometers.

\begin{figure}[ht]
\centerline{\includegraphics[width=0.7\linewidth]{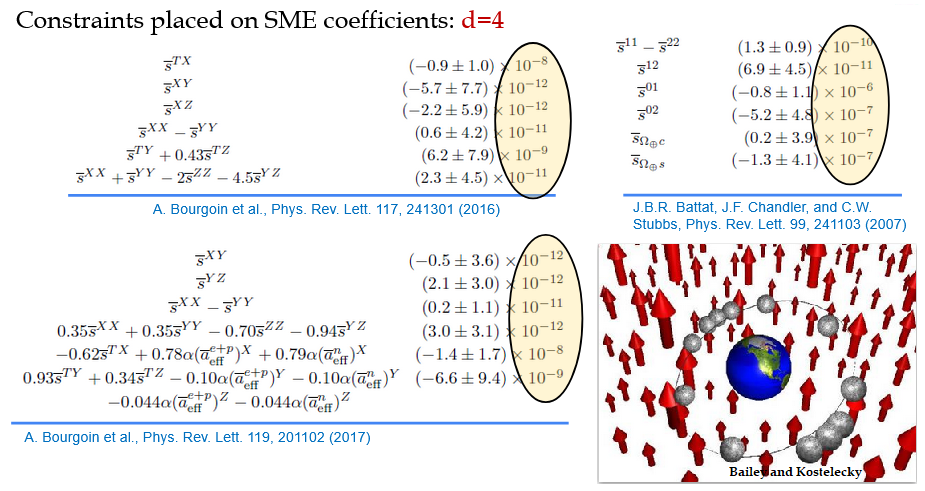}}
\caption[]{Many publications contributed to placing constraints on Lorentz-violating coefficients for gravity terms of mass dimension four from the SME using Lunar Laser Ranging observations \cite{Bourgoin_2016,Bourgoin_2017,james_2007}. The bottom right figure depicts the LV background field with red arrows affecting the moon's orbit around the earth.}
\label{fig:lunarlaserranging}
\end{figure}

When one includes mass dimension four terms that allow for LV from the SME, one can note the first couple extra terms from the SME to the equation of motion between the two bodies,
\begin{equation}
a^J = \frac{G M}{r^3}(\bar{s}^{JK}_t r^K - \frac{3}{2}\bar{s}^{KL}_t\hat{r}^K\hat{r}^L r^J+...),
\end{equation}
where $r^J$ refers to the distance between the Earth and the moon, $G$ is the Gravitational constant, $M$ is their combined mass and $\hat{r}^J=\frac{r^J}{r}$ \cite{Bourgoin_2017}.
Additional terms now contain $\bar{s}^{JK}_t$ coefficients that give rise to the symmetry breaking: it describes an unknown background field that dynamically couples to the two bodies, giving rise to such effects as oscillations in the measured distance. Many publications have provided constraints on these LV coefficients (see figure \ref{fig:lunarlaserranging}).

\subsection{Binary Pulsars}

\begin{figure}[h]
\centerline{\includegraphics[width=0.78\linewidth]{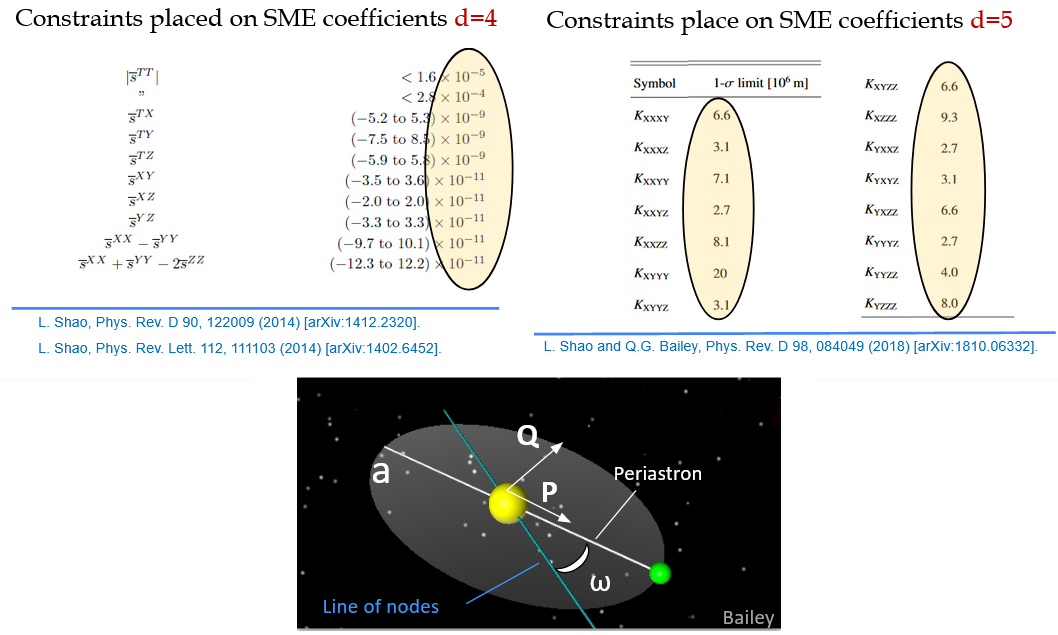}}
\caption[]{Binary Pulsar data has allowed for publications to place constraints on Lorentz-violating coefficients for gravity terms of mass dimension four and five from the SME \cite{lijing_2014prd,lijing_2014prl,lijing_bailey2014}. The bottom figure depicts the binary system and variables representing orbital elements.}
\label{fig:binarypulsars}
\end{figure}

Binary Pulsars are systems of a pulsar with its binary companion, often a white dwarf or neutron star. 
When incorporating mass dimension four and five terms from the SME, the modified Lagrange shows extra terms that allow for spacetime symmetry breaking \cite{shao_bailey_bb,bailey_kost_bb}, 
\begin{align}
    L = &\frac{1}{2}(m_a v_a^2 + m_b v_b^2 )+\frac{Gm_a m_b }{r}(1 + \frac{3}{2}\bar{s}_{00} + \frac{1}{2}\bar{s}_{jk}\hat{n}^j\hat{n}^k)\nonumber\\
    & +\frac{Gm_am_b}{2r}[3\bar{s}_{0j}(v_a^j+v_b^j)+\bar{s}_{0j}\hat{n}^j(v_a^k+v_b^k)\hat{n}^k]-\frac{3Gm_am_b}{2r^2}v_{ab}^j(K_{jklm}\hat{n}^k\hat{n}^j\hat{n}^m-K_{jkkl}\hat{n}^l),
\end{align}
where $m_a$, $m_b$ are the binary masses with their respective velocities $v_a$ and $v_b$. 
The distance between the masses, $r$, has a normal $\hat{n}$ and $G$ is the Gravitational constant. 
The new unknown background coefficients, $\bar{s}_\mn$ and $K_{ijkl}$, allow for spacetime symmetry breaking. 
The resulting effects include secular variations in the orbital elements, modified pulsar timing, and altered orbital velocities and directions. 
Published works have provided constraints to these Lorentz and CPT violating coefficients (see figure \ref{fig:binarypulsars}).

\subsection{3+1, ADM Formulation of the SME, Gravity Sector}

\begin{figure}[h]
\centerline{\includegraphics[width=0.5\linewidth]{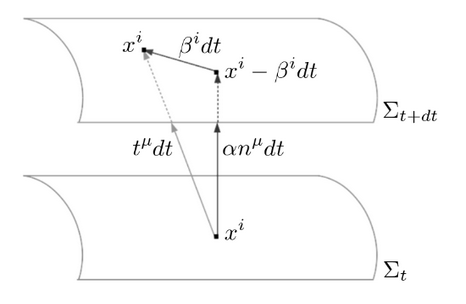}}
\caption[]{In 3+1 formulation, spacetime is foliated into spatial hypersurfaces $\Sigma$ parameterized by time $t$ \cite{oneal2021}.}
\label{fig:3plus1}
\end{figure}

A 3+1 formulation decomposes a 4-dimensional spacetime with the metric $g^{\mu\nu}$, into 3-dimensional spatial hypersurfaces parameterized by time and an associated time-like normal vector $n^{\mu}$ (see figure \ref{fig:3plus1}).
A spatial metric is then $\gamma^{\mu\nu}=g^{\mu\nu}+n^{\mu}n^{\nu}$, $\alpha$ is the Arnowitt-Deser-Minser (ADM) lapse function, $\beta^i$ is a shift vector, and the Riemann tensor is decomposed into a spatial curvature $\mathcal{R}_{\alpha\beta\mu\nu}$ and an extrinsic curvature $K_{\mu\nu}$ \cite{adm,deWitt,mtw,boj}.
Additionally, the spatial covariant derivative is represented with $\mathcal{D}_{\mu}$, and $a_i$ is a spatial acceleration. 

When the lowest order mass dimension four term from the SME is implemented into this process, we arrive at a modified Lagrange,
\begin{align}
\mathcal{L} = \frac {\sqrt{-g}}{2\ka}&[ \cR + K^\ab K_\ab - K^2 - 2 \nabla_\al (n^\al K + a^\al ) \nonumber \\
& + s_\mn \cR^\mn - n^\al n^\be s_\ab (K^\mn K_\mn - K^2 )+...
\end{align}
where $s_{\alpha\beta}$ are the Lorentz-violating coefficients. Then applying the Dirac-Hamiltonian analysis, we arrive with conjugate momentum densities,
\begin{align}
    \Pi_\ga^{ij} &=\frac {\sqrt{\ga}}{2\ka} [\frac{\al^2-s_{00}} {\al^2} \left(K\ga^{ij}-K^{ij}\right) +\frac 1{2\al} \ga^{ij} (\prt_t - \be^k \prt_k) 
    \left( \frac {s_{00}}{\al^2} \right) ], \\
    \Pi_{\be,i} &=0,\\
    \Pi_\al &=\frac{\sqrt{\ga}s_{00}}{\ka \al^3}K,
\end{align}
which provide explicit insights into constraints and degrees of freedom for the system with further investigation. 
The latter set of equations are for a particular case where only the time-component for $s_{\mu\nu}$ is nonzero. 

\begin{figure}[h]
\centerline{\includegraphics[width=0.55\linewidth]{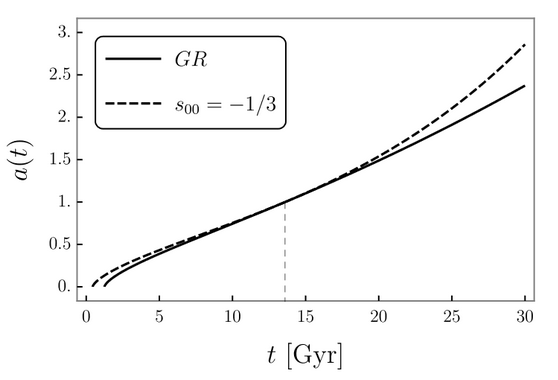}}
\caption[]{Flat FLRW solutions are modified from the LV $s_{00}$ coefficient when compared to that for GR. 
Values of $\Omega_{r0}=0$, $\Omega_{m0}=0.31$, and $\Omega_{\Lambda 0}=0.69$ are assumed. The dashed vertical line represents present day \cite{oneal2021}.}
\label{fig:cosmoEx}
\end{figure}

We apply our work to a special isotropic case in flat Friedmann–Lemaître–Robertson–Walker (FLRW) Cosmology, where our LV coefficient, $s_{00}$, is constant. Here, our Friedmann equations are modified, 
\begin{align}
    \frac {H^2}{H_0^2} =& \Om_{m0} a^{-3} + \Om_{r0} a^{-4\et_r} 
   +\Om_{\La 0} a^{\et_\La} +\Om_{k0} a^{-2}, \nonumber\\ 
   \frac {\ddot{a}}{a H_0^2} =& -\tfrac 12 \Om_{m0} a^{-3} 
-\Om_{r0} \frac {2 (1-s_{00})}{2-s_{00}} a^{-4\et_r} 
+\Om_{\La 0} \frac {2 (1-s_{00})}{2-5s_{00}} a^{\et_\La},
\end{align}
where the $\eta$ within the exponents are fractional expressions for $s_{00}$. 
The interesting outcome shows the density parameter for the cosmological constant is now time dependent, and the rate for the radiation density parameter is altered. The evolution scale factor in this case is compared to that for GR in figure \ref{fig:cosmoEx}, where values of $\Omega_{r0}=0$, $\Omega_{m0}=0.31$, and $\Omega_{\Lambda 0}=0.69$.
Also interesting to note is the modified equation of state we derive form the continuity equation,
\begin{align}
\nabla_\mu[ (T_s)^\mu_{\pt{\mu} 0} &+(T_M)^\mu_{\pt{\mu} 0}]=0\nonumber\\
       \Rightarrow &\dot{\rho}+3\frac{\dot{a}}{a}\frac{2(1 + w -s_{00})} {2 +s_{00}(3w -2)}\rho = 0,
\end{align}
where we do not impose $(T_s)^\mu_{\pt{\mu} 0}=0$ from our LV field \cite{schreck2023,reyes_schreck2021,bonder2022}.

Another motivation for this work was to map the modified gravity model, Ho\v{r}ava-Lifshitz (HL), to terms within the SME. 
HL breaks Lorentz symmetry in the Ultraviolet limit and does not evolve time and space the same way as there are only higher order spatial derivatives. 
The matched expression is then
\begin{equation}
    \cL_{\rm{SME, Match}} = \al \sqrt{\ga} 
\big[ \cR \left(1+ \tfrac 13 s \right) + K^{ij}K_{ij} 
-K^2 (1+\tfrac 12 a_5 )
+ \left(a_{12} + \tfrac 12 a_5 \right) a^i a_i \big],
\end{equation}
where the $a$ coefficients can be substituted to then precisely match HL.

\subsection{Binary Neutron Star Merger: LIGO GW170817}

\begin{figure}[h]
\centerline{\includegraphics[width=0.7\linewidth]{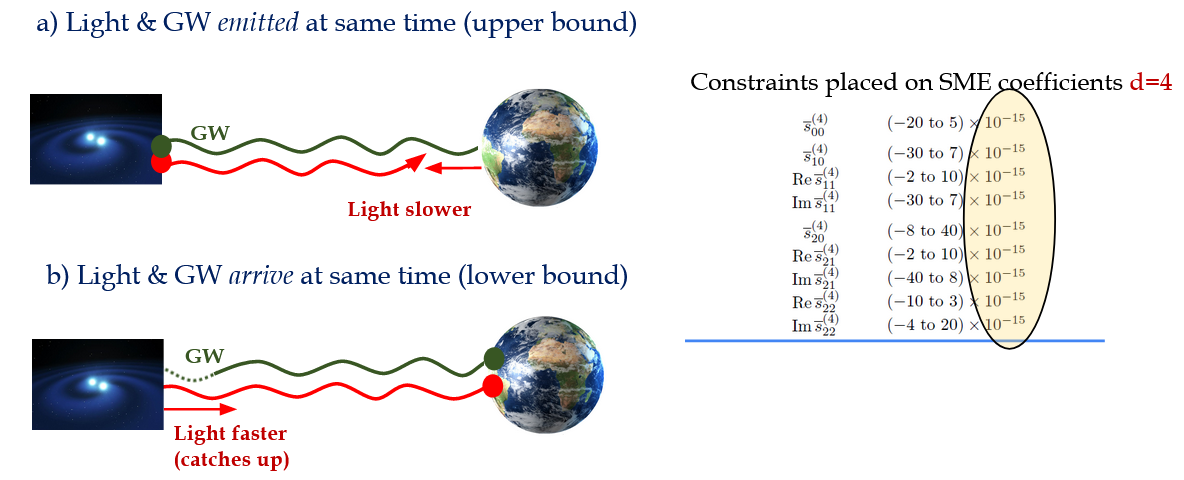}}
\caption[]{Constraints on the possible LV background field are produced by comparing the speeds of the electromagnetic and GW signals \cite{EMvsGW}.}
\label{fig:EMvsGW}
\end{figure}

The binary neutron star merger labeled GW170817 provided the first event to have both a gravitational wave (GW) and electromagnetic signal. 
In this way, their speeds can be compared by measured the arrival times of the two types of signals. When these waves propagate through spacetime, a possible LV background field would potentially alter their speed from the expected $c$. 
Lower and upper bounds can be placed on a fractional difference that depends on whether the signals are emitted at the same time and arrive at different moments at the detectors, or instead arrive at the same time after being emitted at different moments at the source, $-3\,\cdot\,10^{-15}\leq \frac{\Delta \nu}{\nu_{EM}}\leq +7\,\cdot\,10^{-16}$  \cite{EMvsGW}. 
Furthermore, constraints are then placed on SME coefficients, one at a time, for mass dimension four (see figure \ref{fig:EMvsGW}).

\subsection{Birefringence in Gravitational Waves}

Searching for Lorentz and CPT symmetry breaking during the propagation of gravitational waves is done for SME terms of mass dimension five. 
Gravitational waves are considered small perturbations in Minkowski spacetime, represented by the metric as $g_{\mu\nu}=\eta_{\mu\nu}+h_{\mu\nu}$ where $h_{\mu\nu}<<\eta_{\mu\nu}$.
Assuming linear GR gauge symmetry and retaining only terms second order in $h_{\mu\nu}$, our Lagrange will have the first term for linearized GR and additional terms that allow for symmetry breaking \cite{km09,mewes2019}:
\begin{equation}
     \cL = \frac{1}{8\ka} \ep^{\mu\rh\al\ka}\ep^{\nu\si\be\la}\eta_{\ka\la}h_{\mu\nu}\prt_{\al}\prt_{\be}h_{\rh\si} 
        +\frac{1}{8\ka} h_{\mu\nu}(\hat{s}^{\mu\rh\nu\si}+\hat{q}^{\mu\rh\nu\si}
        +\hat{k}^{\mu\rh\nu\si})h_{\rh\si}.
        \label{bireL}
\end{equation}
The $\hat{s}^{\mu\rh\nu\si}$, $\hat{q}^{\mu\rh\nu\si}$, and $\hat{k}^{\mu\rh\nu\si}$ are field operators than can be written as, 
\begin{align}
    \hat{s}^{\mu\rh\nu\si}=&s^{(d)\mu\rh\ep_1\nu\si\ep_2...\ep_{d-2}}\prt_{\ep_1}...\prt_{\ep_{d-2}}, \nonumber\\
    \hat{q}^{\mu\rh\nu\si}=&q^{(d)\mu\rho\ep_1\nu\ep_2\si\ep_3...\ep_{d-2}}\prt_{\ep_1}...\prt_{\ep_{d-2}} \nonumber\\
    \hat{k}^{\mu\nu\rh\si}=&k^{(d)\mu\ep_1\nu\ep_2\rh\ep_3\si\ep_4...\ep_{d-2}}\prt_{\ep_1}...\prt_{\ep_{d-2}},
\end{align}
where coupled to the derivatives are assumed to be small, unknown background fields referred to as coefficients or parameters. 
These coefficients are constrained experimentally.
There are three labeled filed operators due to being categorized as either CPT odd or even and what mass dimension they contain. 

\begin{figure}[h]
\centerline{\includegraphics[width=0.7\linewidth]{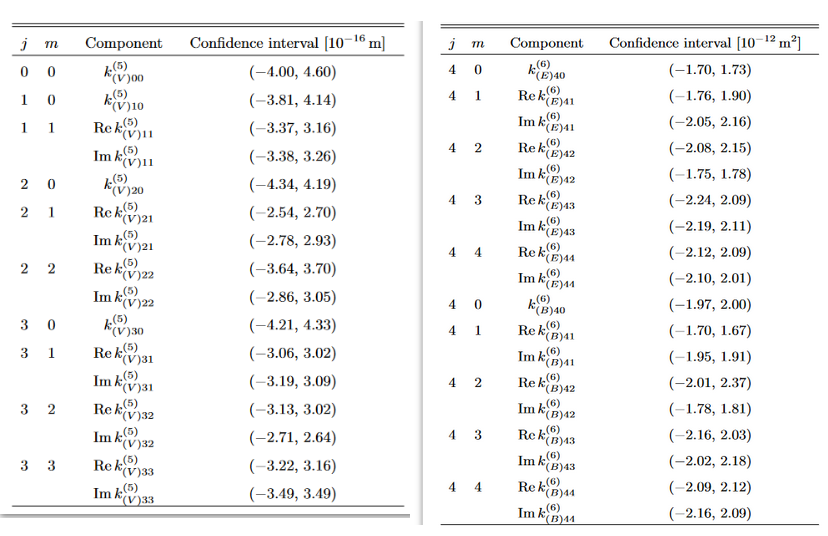}}
\caption[]{Constraints on the possible symmetry-breaking background field are produced by noting the possible extra delay in arrival times between the two GW polarizations are limited to detector noise. 
The left table shows $68\%$ confidence intervals for the 16 independent components from the mass dimension five SME term; the right table shows the same but for mass dimension six terms \cite{wang2021}.}
\label{fig:lijing}
\end{figure}

Initial constraints were done by noting the time delay between the two, linearly independent gravitational wave modes could have additional delay from LV and CPT violating effects, $\Delta t$, which when comparing their peak frequencies, f, would be limited by the detector noise $\rho$, i.e., $|\Delta t|\leq \frac{1}{\rho f}$ \cite{lijing2020,wang2021} (see figure \ref{fig:lijing}).

Deriving the equations of motion from Eq. \eqref{bireL} and transforming into frequency-momentum space, the dispersion relation, 
\begin{equation}
    \omega = |\vec p| \, 
    \left( 1-\zeta^0 \pm |\vec{\ze}| \right),
    \label{dispEq}
\end{equation}
now includes zeta terms that are expressions of the previous symmetry-breaking coefficients. 
The $\pm$ term allows for the effect of birefringence, where the dynamical coupling of the unknown LV and CPT violating field to the two linearly independent polarizations results in a possible extra difference in their arrival times. 
The effect can be seen as a phase shift in the GWs, $\psi_{\pm}=-\delta \pm \beta$, which after including cosmological redshift and using celestial coordinates, we find modified expressions for the two GW polarizations, 
\begin{align}
    h_{(+)} =& e^{i\de} (\cos \be - i \sin \vth \cos \vph \sin \be )\, h^{LI}_{(+)}  - e^{i\delta}\sin \be (\cos \vth + i \sin \vth \sin \vph )  \, h^{LI}_{(\times)}, \nonumber\\
     h_{(\times)}=& e^{i\de} (\cos \be +i \sin \vth \cos \vph \sin \be )\, h^{LI}_{(\times)} + e^{i\de}\sin \beta(\cos \vth - i \sin \vth \sin \vph )  \, h^{LI}_{(+)}.
\end{align}
The $h^{LI}_{(+)}$ and $h^{LI}_{(\times)}$ are the GR polarizations with $\vth$ and $\vph$ sufficiently refer to the sky localization angles for the source.

\begin{figure}[h]
\centerline{\includegraphics[width=0.78\linewidth]{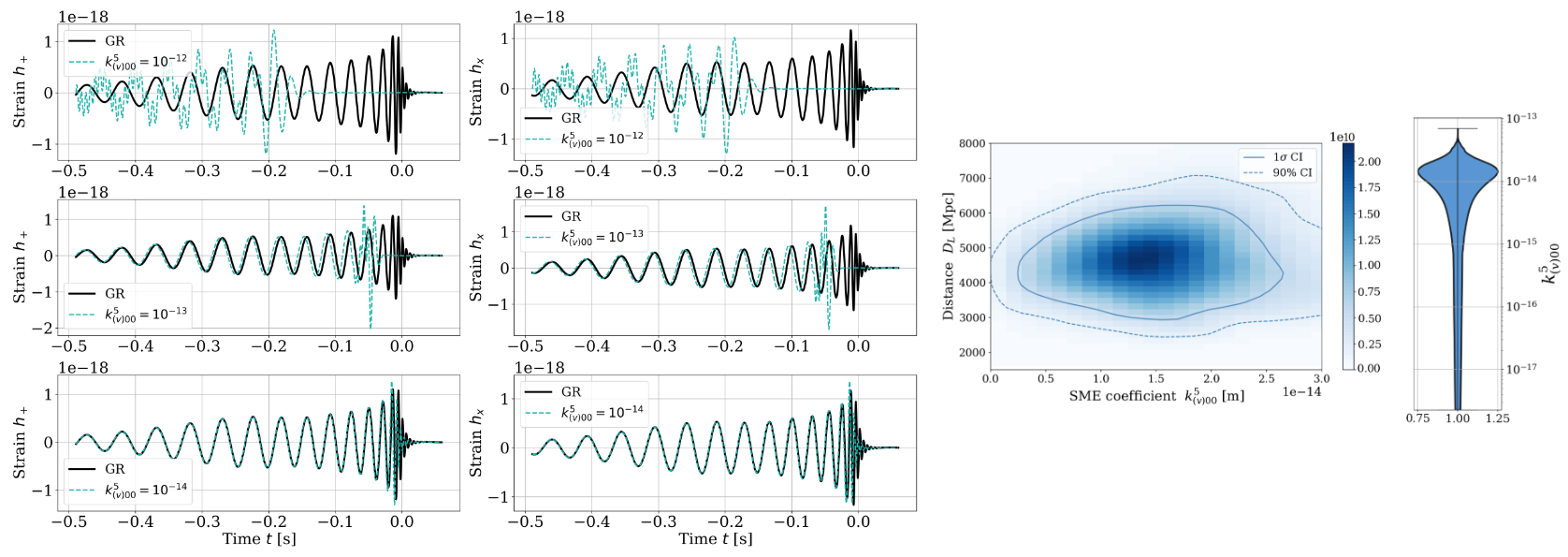}}
\caption[]{All are sensitivity studies for implementation of symmetry-breaking modifications into LALSuite. The left waveforms have varying $k^{(5)}_{(V)00}$ values for a simulated system of non-spinning binary black holes with $m1=m2=50$ solar masses at a luminosity distance of $4$Gpc. The two figures on the right show posterior probability density results for the same system but for $5$Gpc. The first of the two shows the 1$\sigma$ and $90\%$ credible intervals between the degenerate $k^{(5)}_{(V)00}$ and $D_L$ parameters. The far right is the marginalized  posterior probability for $k^{(5)}_{(V)00}$.}
\label{fig:sensistudies}
\end{figure}

\begin{figure}[h]
\centerline{\includegraphics[width=0.78\linewidth]{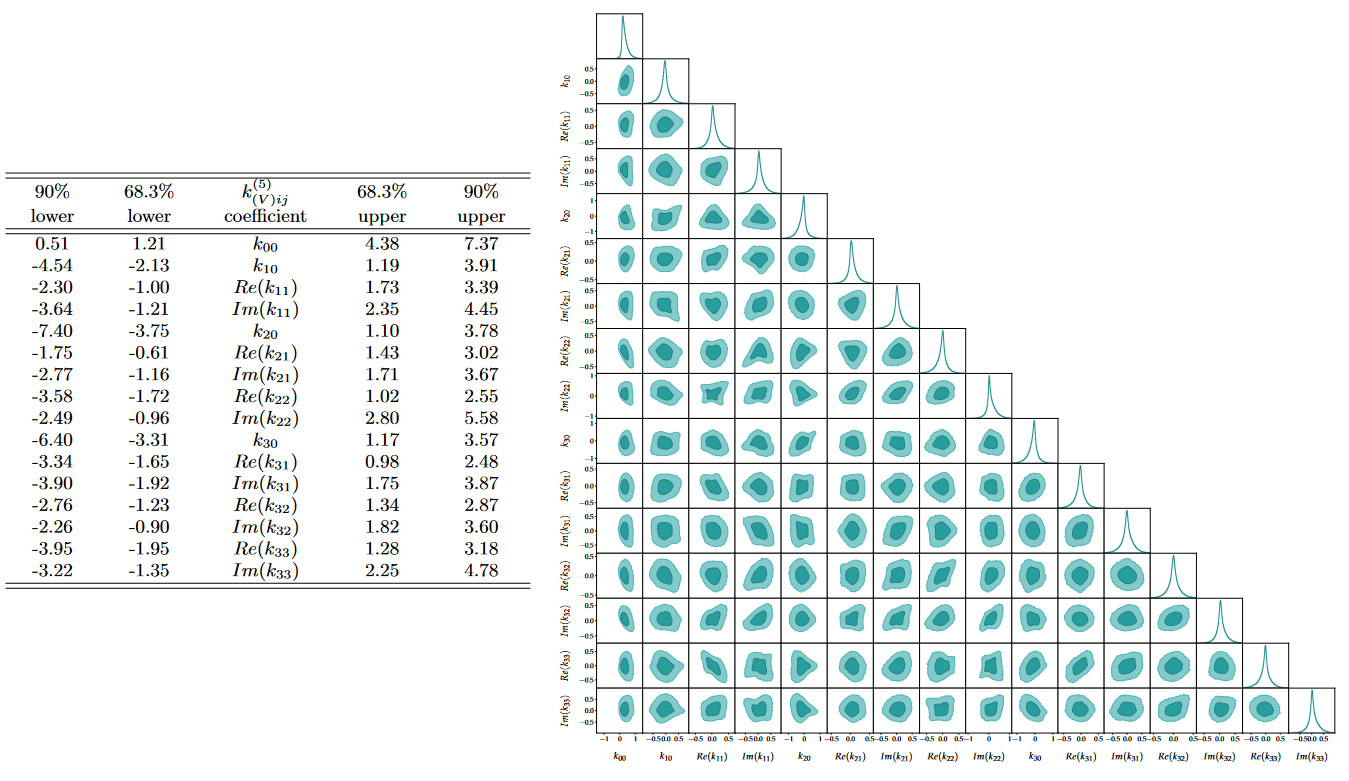}}
\caption[]{Credible intervals on the $k^{(5)}_{(V)ij}$ symmetry-breaking coefficients are displayed in the left table. Posterior probabilities for the same coefficients are shown on the right. Darker shaded areas are the $68.3\%$ credible intervals, the light blue $90\%$. }
\label{fig:bireResults}
\end{figure}

This theory is implemented into \textit{LIGO Scientific Collaboration Algorithm Library Suite} (LALSuite). Modifications of the GW signals derived are implemented into LALSuite's package, LALSimulation, focusing on the simplest coefficients of mass dimension five. Sensitivity studies were conducted, providing reasonable values for parameters, e.g., $m1=m2=50$ solar masses for the binaries coalescing and a luminosity distance of $4$Gpc. Different magnitudes of the symmetry breaking coefficient, $k^{(5)}_{(V)00}$, were explored and were noted to be of acceptable values \cite{oneal2021,haegel2023} (see figure \ref{fig:sensistudies}).
The LALInference package of LALSutie is also modified, where the GW strain is directly analysed via Bayesian inference of posterior probabilities of the GW source parameters. Forty-five events from the GWTC-3 Ligo-Virgo-Kagra (LVK) catalogue are analysed with false-alarm rates (FAR) of less than $1/$year from O1 and O2 events, and FAR $<10^{-13}$/year from O3a and O3b events. Individual constraints were combined post-processing to measure the 16 anisotropic coefficients jointly using the SVD Method. Sensitivity studies are displayed in figure 8; published constraints are displayed in figure \ref{fig:bireResults}. 

\section{Summary}

The SME framework provides a generic way to test the breaking of underlying symmetries built into the theories of GR and the SM. These symmetries include Lorentz and CPT. 
There have been many constraints from decades of previous experiments and theoretical developments that have been placed on such symmetry-breaking coefficients within this framework. Specific modified physics models are also mapped to terms within the SME, like that of Ho\v{r}ava-Liftshitz gravity. 
Yet still, there is much phenomenology left to investigate within the many terms of this framework from all sectors of physics and much collaborative work to be done.

\section*{Acknowledgments}

KOA gratefully acknowledges the $57^{th}$ Rencontres de Moriond organizers for the invitation and opportunity to participate. 
Support for this work and conference travel was provided by the Department of Physics and Astronomy at Embry-Riddle Aeronautical University, Prescott, AZ in the United States and the National Science Foundation grant number 2207734.



\section*{References}

\begin{thebibliography}{99}

\bibitem{kost_2004}V. A. Kosteleck{\'{y}}, \Journal{\PRD}{69}{10}{2004}.

\bibitem{coll_kost_1998}D. Colladay and V. A. Kosteleck{\'{y}}, \Journal{\PRD}{58}{11}{1998}.

\bibitem{data_tables}V. A. Kosteleck{\'{y}
} and Neil Russell, \Journal{\textit{Reviews of Modern Physics}}{83}{1}{2011}.

\bibitem{bk06}S. M. Carroll, \Journal{\PRL}{87}{14}{2001}.

\bibitem{bc2018}Q. Bailey and C. Lane, \Journal{\textit{Symmetry}}{10}{10}{2018}.

\bibitem{jhf2021}J. Overduin et al, \Journal{\textit{Galaxies}}{9}{2}{2021}.

\bibitem{LLR_1962}L. D. Smullin and G. Fuicco, \Journal{\textit{Nature}}{192}{1267}{1962}.

\bibitem{Bourgoin_2016}A. Bourgoin et al, \Journal{\PRL}{117}{24}{2016}.

\bibitem{Bourgoin_2017}A. Bourgoin et al, \Journal{\PRL}{119}{20}{2017}.

\bibitem{james_2007}J. B. R. Brattat et al, \Journal{\PRL}{99}{24}{2007}.

\bibitem{shao_bailey_bb}L. Shao and Q. G. Bailey, \Journal{\PRD}{98}{8}{2018}.

\bibitem{bailey_kost_bb}{Q. G. Bailey and V. A. Kosteleck{\'{y}
}}, \Journal{\PRD}{74}{4}{2006}.

\bibitem{ADM_SME}K. O'Neal-Ault and Q. G. Bailey and N. A. Nilsson, \Journal{\PRD}{103}{4}{2021}.

\bibitem{EMvsGW}B. P. Abbott et al, \Journal{\textit{The Astrophysical Journal}}{848}{2}{2017}.

\bibitem{km09}Kosteleck\'y, V.A. and Mewes, M.,
\Journal{\PRD}{80}{1}{2009}

\bibitem{mewes2019}M Mewes, \Journal{\PRD}{99}{10}{2019}.

\bibitem{lijing2020}L. Shao, \Journal{\PRD}{101}{10}{2020}.

\bibitem{wang2021}Z. Wang, \Journal{\textit{The Astrophysical Journal}}{921}{2}{2021}.

\bibitem{lijing_2014prd}L. Shao,
\Journal{\PRD}{90}{12}{2014}.

\bibitem{lijing_2014prl}L. Shao,
\Journal{\PRL}{112}{11}{2014}.

\bibitem{lijing_bailey2014}L. Shao and Q. G. Bailey,
\Journal{\PRD}{98}{8}{2018}.

\bibitem{adm}R. Arnowitt et al,
Phys.\ Rev.\ {\bf 116}, 1322 (1959).

\bibitem{deWitt}B.S. DeWitt, 
Phys.\ Rev.\ {\bf 160}, 1113 (1967).

\bibitem{mtw}C.S. Misner et al, \textit{Gravitation} 
(W.H.\ Freeman and Company, New York, 1973).

\bibitem{boj}M. Bojowald,
{\it Canonical Gravity and Applications},
(Cambridge University Press, 2011).


\bibitem{schreck2023}M. Schreck, {\textit{The Ninth Meeting on CPT and Lorentz Symmetry}}{bf eprint: 2301.13008}{2023}.

\bibitem{reyes_schreck2021}C. M. Reyes and M. Schreck, \Journal{\PRD}{104}{12}{2021}.

\bibitem{bonder2022}Y. Yuri, {\textit{The Ninth Meeting on CPT and Lorentz Symmetry}}{bf eprint: 2206.10759}(2022).



\bibitem{oneal2021}K. O'Neal-Ault et al, \Journal{\textit{Universe}}{7}{10}{2021}.

\bibitem{haegel2023}L. Haegel et al, \Journal{\PRD}{107}{6}{2023}.



\end{thebibliography}
\end{document}